\documentclass[11pt,twocolumn]{article}
\usepackage{amsmath}
\usepackage{graphicx}
\usepackage{tabularx}
\usepackage{epstopdf}
\usepackage{color}
\usepackage[a4paper,left=2cm,right=2cm,top=2cm,bottom=2cm]{geometry}

\title{Element Decoupling of 7T Dipole Body Arrays \\by EBG Metasurface Structures: Experimental Verification}

\author{
  Anna~A.~Hurshkainen, Tatyana~A.~Derzhavskaya, Stanislav~B.~Glybovski, Irina~V.~Melchakova\\
  Department of Nanophotonics and Metamaterials, \\
  ITMO University, 197101 St. Petersburg, Russia
  \\
  \texttt{s.glybovski@metalab.ifmo.ru}
  \and
  Alexander~J.E.~Raaijmakers, Ingmar~J.~Voogt, Cornelis~A.T.~van~den~Berg\\
  University Medical Center Utrecht, 3584 CX, Netherlands\\
  \texttt{a.raaijmakers@umcutrecht.nl}
}

\date{\today} 
  
\begin {document}  
  \maketitle
     
\begin{abstract}
Metasurfaces are artificial electromagnetic boundaries or interfaces usually implemented as two-dimensional periodic structures with subwavelength periodicity and engineered properties of constituent unit cells. The electromagnetic bandgap (EBG) effect in metasurfaces prevents all surface modes from propagating in a certain frequency band. While metasurfaces provide a number of important applications in microwave antennas and antenna arrays, their features are also highly suitable for MRI applications.
In this work we manufacture and experimentally study finite samples based on mushroom-type EBG metasurfaces and employ them for suppression of inter-element coupling in dipole transmit coil arrays for body imaging at 7T.
We show experimentally that employment of the samples EBG leads to reduction of coupling between adjacent closely-spaced dipole antenna elements of a 7T transmit/receive body array, which reduces scattering losses in neighboring channels and thereby improves the B1+ efficiency. 
The setup consists of two fractionated dipole antennas previously designed by the authors for body imaging at 7 Tesla. These are placed on top of a body-mimicking phantom and equipped with the manufactured finite-size sample of the metasurface tuned to have EBG properties at the Larmor frequency of 298 MHz.
To improve the detection range of the B1+ field distribution of the top elements, four additional elements were positioned along the bottom side of the phantom. Scattering matrix measurements show that coupling between the two top elements is indeed reduced while the measurements performed on a 7T MRI machine confirm the array's B1+ efficiency improvement due to 
reduced scattering losses.
This study provides a tool for the decoupling of dipole antennas in ultrahigh field transmit arrays, possibly resulting in denser element placement and/or larger subject-element spacing. 
\end{abstract}

\section{Introduction}
For ultrahigh field imaging, the small wavelength in tissue renders the commonly used birdcage body coil obsolete due to destructive interferences. These interferences cannot be avoided but can be controlled using a multi-transmit setup where a transmit coil array is being driven with a given number of independent channels, with optimized phases and amplitudes. To arrive at sufficient efficiency, these transmit coil arrays are usually surface arrays, i.e. the elements are placed directly on the subject. The minimal element-subject spacing ensures optimal efficiency of the elements, which is needed because of the limited available transmit power. For the same reason, inter-element coupling of the elements should preferably be minimal to avoid that RF power gets lost by scattering into the other channels and ending up in the dummy load of the circulators. 

Various types of transmit array elements have been presented by various institutes. Some of these coil array element types have ways to reduce or avoid coupling. For loop coils, this can be achieved by overlap of neighboring elements \cite{surfaceloop}; for microstrip TEM arrays, the elements can be decoupled by capacitors between the ground planes \cite{TEMdec}. Next to these, also dipole antennas have demonstrated to be very suitable transmit coil array elements at ultrahigh frequencies \cite{Alexander,AlexMRM20015}. However, decoupling of dipole antennas is difficult or impossible with methods conventional for MRI. Although the applied dipole antennas have inherently low coupling, this coupling still imposes minimum inter-element spacing requirements and, therefore, imposes a maximum on the channels that can be positioned along the body circumference. Decoupling methods would enable the use of a larger number of elements, to position the elements more closely together around the imaging target (e.g. heart) or to increase the element-subject distance with beneficial impact on the SAR levels. 

One approach coming from the antenna array theory is to use passive matching decoupling circuits to feed array elements. The parameters of the 2N-port decoupling circuit can be synthesized based on the known impedance matrix coefficients of a N-port produced by N interacting antenna array elements \cite{pdn}. Since the implementation of the required matching-decoupling circuit is not unique, there are approaches to help in the circuit synthesis based on an explicit criteria of a maximized efficiency,e.g. the Eigen-Analysis theory \cite{Hein}. The common disadvantage of matching-decoupling circuits is their high insertion losses and  most of all  they only work for one given phase and amplitude setting while MRI transmit arrays need degrees of freedom in the drive settings to ensure constructive interference in the imaging target for a wide range of subjects and imaging regions. In some other techniques active feeding networks are used to compensate the signals induced in each channel due to coupling with the other ones, e.g. the Cartesian feedback method \cite{cartesian}. However, this does not make the system more power efficient because scattered power needs to be compensated by additional forward power on each channel. 

One possible approach would be to use passive inter-element dipoles \cite{dipoledec}. A similar approach  with passive loops has been applied before also for loop coil arrays \cite{passiveloop}. Recently a novel technique based on the magnetic wall distributed filters, which act as passive decoupling structures has been proposed and experimentally realized in \cite{magwall1}. In this technique the elements of a loop coil head array are separated by linear periodic rows comprising multiple magnetically polarizable and small resonant unit cells. The shape of such inclusions was inspired by designs of frequency-selective surfaces and implemented by the authors using PCB technology. For the setup having three loops operating at 7T the method provides isolation of -22 dB for inter-element separation of
7 mm. The same filters have been used for decoupling of a 7T transceive coil array in \cite{magwall2}. However, despite of such a strong isolation, some decrease  of the transmit efficiency by 16\% was observed in the presence of these filters.

Another approach presented in this paper is the use of resonant metasurfaces. 
Metasurfaces are usually periodic two-dimensional arrays of structured inclusions with sub-wavelength dimensions and periodicity. Metasurfaces effectively behave as thin continuous sheets with respect to the excitation
electromagnetic field. In this behavior metasurfaces exhibit certain required electromagnetic
properties achieved through engineering of the inclusion microstructure. Metasurfaces though
being electrically thin provide a number of advantageous functions of control over electromagnetic waves \cite{Holloway,Capasso}, e.g. wave focusing \cite{focusing}, polarization conversion \cite{polarization} or manipulation with localized surface waves \cite{manipulation}. 

Decoupling of various microwave antennas by means of resonant metasurfaces is a well elaborated topic. For this purpose a specific type of metasurfaces is usually employed called the High
Impedance Surfaces (HIS). One of the effects of such structures is propagation suppression of any type of surface modes within the so-called electromagnetic bandgap (EBG) \cite{capolino2009theory} first shown in \cite{Sievenpiper}. 
When located nearby a surface with special electromagnetic properties an antenna can actively excite surface modes rather than modes of free space. The EBG regime means that in a certain frequency range all surface modes become evanescent and cannot propagate along the structure though still being excited by the antenna.  In this regime the field created by the antenna just above the surface fastly reduces with a distance as the surface mode fields decay exponentially. Therefore, the EBG effect can be used to suppress the link between two antennas both situated close enough to the surface. 
The applications of high impedance metasurfaces in the microwave range are elimination of strong mutual coupling between antenna array elements \cite{Yan03,Bro05} and suppression of edge diffraction effects on antenna ground planes \cite{Bag08}. In fact, unit cells of any metasurface have finite dimensions, so that it is possible to contain a limited number of unit cells between two antennas. However, as demonstrated in \cite{Yan03}, noticeable coupling suppression can be achieved even with finite EBG metasurface samples having electrically small dimensions. 

Another useful feature of HIS is the artificial magnetic conductor (AMC) property \cite{Sievenpiper,capolino2009theory}. In this regime the infinite structure behaves as a magnetic reflector, which creates in-phase reflection instead of out-of-phase reflection provided by conventional (electric) reflectors made of metal. This allows horizontal dipoles to be mounted close to the AMC surface producing highly-efficient and low-profile directed antennas without destructive interference. 

The first proposal of using HIS for improvement of 7T MRI coils can be found in \cite{Saleh12low}, where also an appropriate miniaturized unit-cell design was discussed. In \cite{Saleh12}  it was proposed to use stacked offset HIS as artificial substrates improving B1+ efficiency of a strip-line coil at 7T instead of conventional metallic ground planes. It was shown by numerical simulations that the B1+ field can be enhanced all over the homogeneous phantom by about 47\%  due to suppression of destructive interference by a HIS sample. Also HIS substrates were proposed to improve the E/H field ratio of strip-line coils \cite{Saleh13}.

In this work we experimentally realize the high-impedance stacked metasurface of the so-
called mushroom design \cite{Sievenpiper,capolino2009theory} having an electromagnetic bandgap at nearly 298 MHz and employ it for decoupling of adjacent electric dipole antennas of a 7T body array. By means of measurements on a 7T laboratory MRI machine we for the first time observe improvement of B1+ efficiency due to the presence of the finite-size EBG metasurface sample in vicinity of  two of six array elements (fractionated dipole antennas \cite{AlexMRM20015} i.e. segmented dipole antennas with meanders to reduce SAR  levels \cite{Alexander}). Additional measurements of S-parameters and numerical simulations confirm that the observed efficiency improvement is explained by reduction of energy dissipation through the coupling suppression between the corresponding dipoles as well as the constructive interference effect of the finite-size metasurface.

This paper is organized as follows. In Section \ref{Design} we characterize the employed EBG metasurface design and show the results of its numerical optimization for the 7T MRI application, in Section \ref{Exp} the methods of experimental study of the array are described including measurements of S-parameters and measurements on a laboratory MRI machine, while in Section \ref{Conclude} we discuss the results and make a conclusion.

\section{Design of EBG Structure}\label{Design}

In most of MRI applications including 7T systems the distance between adjacent elements of body arrays is strongly limited. For example, the element separation of a eight-element body array at 7T falls into the range 0.01--0.1 of the wavelength in free space. To realize an EBG metasurface, which efficiently isolates such closely-separated dipole antennas by suppression of surface wave propagation, one should achieve extremely small periodicity and the unit-cell dimensions. The metasurface sample must contain at least several unit cells on the way of the surface wave travelling between the dipoles \cite{Saleh12}. On another hand each unit cell must be resonant in order to sufficiently eliminate the effect of propagation and make all surface mode evanescent. These conditions require the unit cells to be strongly miniaturized resonators with the resonant frequency of around 298 MHz and the dimensions of the order of 10 mm. This can be realized by introducing large capacitance and/or large inductance of the unit cells.   

\begin{figure}[!h]
  \centering
  \includegraphics[width=0.7\linewidth]{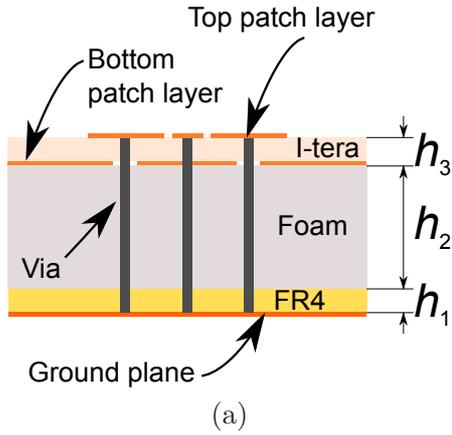}\\(a)\\~~\\
  \includegraphics[width=0.7\linewidth]{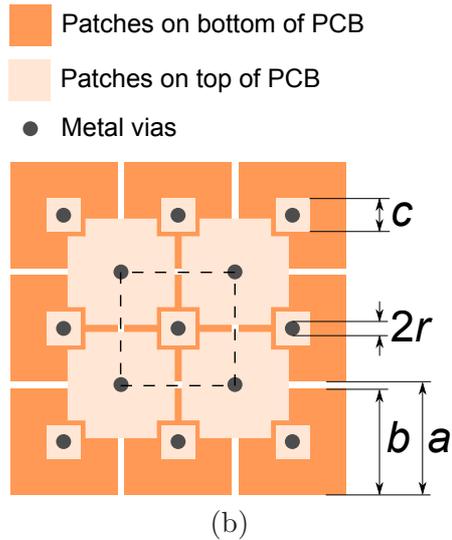}\\(b)
  \caption{Unit cell of the stacked mushroom-type EBG metasurface with overlapping patches and additional thick foam layer: (a) side view, (b) top view}
  \label{Fig1}
\end{figure}

We employ the multilayer \textit{mushroom}-type EBG metasurface originally proposed in \cite{Sievenpiper} for miniaturization of resonant unit cells. The advantage of this metasurface is its isotropy and insensitivity of the EBG properties to polarization, i.e. the structure in its electromagnetic bandgap regime suppresses surface waves with arbitrary polarization traveling in any in-plane direction. In this work we implement the structure stacking two printed-circuit boards: the bottom one with FR4 substrate realizing the common ground plane of unit cells and the top one with I-tera substrate supporting overlapping square metal patches on the opposite sides of a dielectric substrate, as shown in Figure \ref{Fig1}. 
Since the upper substrate thickness $h_3$ is only 0.5 mm and the permittivity is $\varepsilon_{3,r}=3.45$, the overlapping patches introduce a high capacitance. The dielectric loss tangent of the patch substrate is 0.0031 only, which minimizes dissipation losses in the regions of high electric field, i.e. between the patches. For the bottom PCB the substrate material with the thickness of $h_2=1.5$ mm does not affect the properties of unit cells, since the layer appears in the zero of electric field distribution (near the ground plane). To increase the unit cell inductance and further reduce the resonant frequency the overall structure's thickness is increased by adding a foam layer of the acceptable thickness of $h_2$ and $\varepsilon_{2,r}=1$ between the two PCBs. All patches are then connected to the ground plane by metal vias, which are realized as vertical copper wires of radius $r=0.6$ mm going through holes in the foam layer, as shown in Figure \ref{Fig1}(a).

For the experiment we fabricated the EBG sample with 3 unit cells in transverse direction to the dipole's axes and 15 unit cells in the longitudinal direction. The patches on the bottom layer of the upper PCB are soldered to the vias on the top interface of the structure, which is possible because of additional soldering square plates with the side $c=3$ mm shown in Figure \ref{Fig1}(b).

\begin{figure}
  \centering
  \includegraphics[width=1\linewidth]{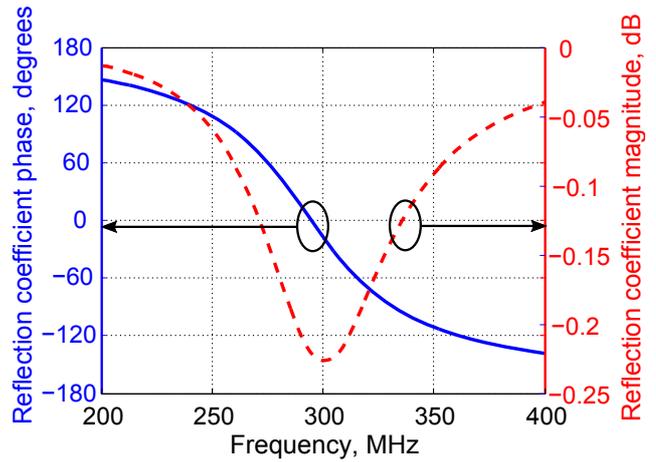}
  \caption{Magnitude and phase of the plane-wave reflection coefficient for the infinite metasurface composed of optimized unit cells (CST Simulation)}
  \label{Fig2}
\end{figure}

\begin{figure}
  \centering
  \includegraphics[width=1\linewidth]{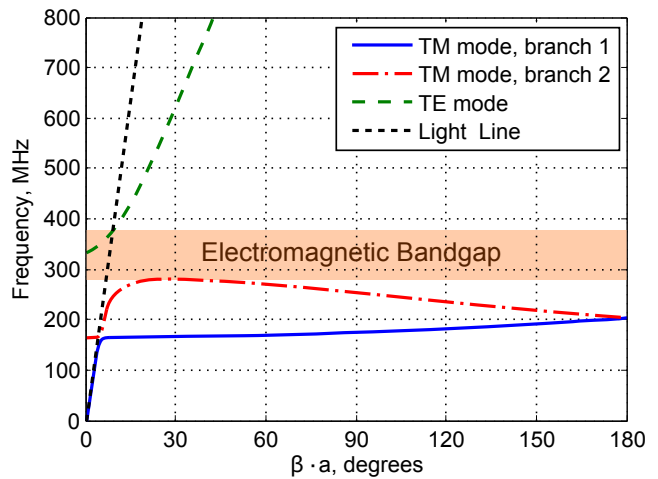}
  \caption{Surface-wave dispersion diagram of the infinite metasurface composed of optimized unit cells (CST Simulation)}
  \label{Fig3}
\end{figure}

The dimensions of patches $b=19.5$ mm, their period $a=20$ and the foam layer thickness $h_2=37$ mm were chosen to ensure EBG and AMC properties at 298 MHz. For the optimization of the unit-cell parameters we employed the commercial simulation software CST Microwave Studio 2014. We employed Floquet ports together with two pairs of periodic boundary conditions to simulate a singe unit cell in an infinite array. The reflection coefficient's magnitude and phase of the the optimized metasurface calculated in the frequency range 200--400 MHz are presented in Figure \ref{Fig2}.
Also the dispersion diagram of the infinite EBG metasurface with optimized unit cells was computed using Eigenmode Solver. The results are presented in Figure \ref{Fig3} for the frequency dependence of the propagation factors of TM- and TE-polarized surface waves.
From Figure \ref{Fig2} it is seen that the resonance providing magnetic wall properties (zero reflection coefficient's phase) takes place at about 300 MHz. That means that the metasurface at least partially demonstrates constructive interference with respect to a horizontal dipole antenna located nearby the interface \cite{Sievenpiper}. The reflection coefficient's magnitude at the resonance reaches -0.22 dB, which confirms low dissipation losses within the metasurface despite of the small electrical dimensions of unit cells (0.02 of the wavelength).

Inspecting Figure \ref{Fig3} one can conclude that for the optimized unit-cell geometry the structure exhibits EBG properties in the range 290--380 MHz, where neither TM- nor TE-polarized surface modes can propagate (there are no real solutions for the propagation factor $\beta$ at these frequencies). The EBG mode is highlighted in Figure \ref{Fig3}.

\section{Experiment}\label{Exp}

\begin{figure} [!h]
  \centering
  \includegraphics[width=0.8\linewidth]{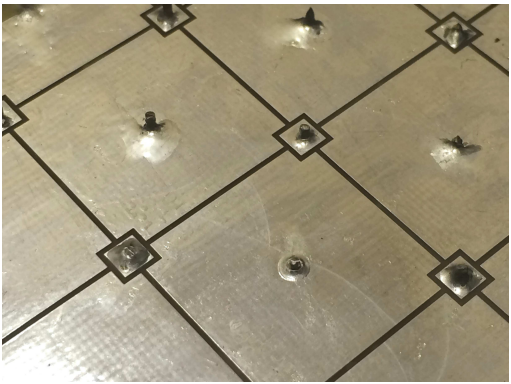}
  \caption{Photograph of the top interface of the manufactured EBG metasurface}
  \label{FigPhoto}
\end{figure}

In this section we describe a proof-of-principle experiment which shows that the presence of the optimized metasurface sample improves B1+ efficiency of a transceive body array with fractionated 300-mm-length dipole elements. The top layer of patches of the manufactured metasurface is depicted in Figure \ref{FigPhoto}. 

\begin{figure} [!h]
  \centering
  \includegraphics[width=0.9\linewidth]{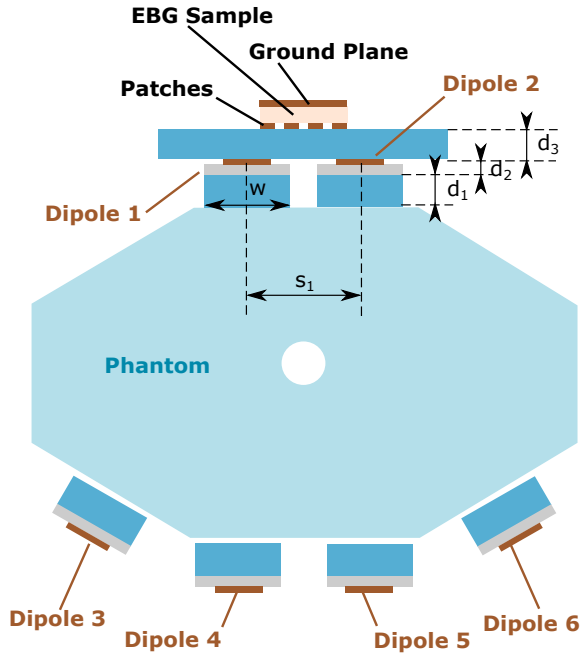}
  \caption{Experimental setup with six-element body array with dipole elements surrounding the homogeneous phantom with the circular hole at the center. Two of six antennas on top are equipped with the metasurface sample}
  \label{FigSetup}
\end{figure}

For the experimental investigation we consider six-element array of identical fractionated dipoles \cite{AlexMRM20015}, surrounding a homogeneous phantom (water solution of Ethylene-glycol-NaCl, 34 g/l with the relative permittivity of 45 and conductivity of 0.25 S/m). The phantom has a hole in the center woth the diameter of 34 mm to place a magnetic field probe inside. The arrangement of the dipoles is illustrated in Figure \ref{FigSetup}. Each dipole consists of a copper trace printed on a 1.5 mm FR4 substrate having the dimensions of $30\times 308$ mm, which is, in turn, located over the plexi-glass plate of the thickness $d_2=18$ mm. The antennas are separated from the phantom's surface by foam layers of the thickness $d_1=25$ mm. The EBG metasurface sample is located above the two antennas on top of the phantom, i.e. Dipoles 1 and 2 (see Figure \ref{FigSetup}). The sample's boundary with patches is directed towards the antennas. There is also a foam layer of the thickness $d_3=45$ mm separating the sample from Dipoles 1 and 2. This thickness $d_3$ is chosen experimentally in order to maximize B1+ field inside a phantom, i.e. to tune the condition of the constructive interference between the antenna's field and the field scattered by the sample. The possibility to reach this condition with the manufactured sample is provided by the AMC effect of the corresponding infinite metasurface around 300 MHz. On another hand, since the sample if finite and coupled to the antennas, a small phase shift of the scattered field occurs, which can be compensated by the proper choice of $d_3$. Therefore, the sample being located close to the antennas and demonstrating decoupling capabilities also causes constructive interference.

The other four antennas are located on the opposite side of the phantom and only serve to improve the receive performance for B1+ mapping. 
We note that in such an arrangement Dipoles 1 and 2 are only $s_1=30$ mm away from each other ($s_1$ is the distance between the axes of the dipoles), and therefore, suffer from high mutual coupling rather than coupling to other antennas. The aim of the experiment is to study the effect of the metasurface sample, which isolates the two neighboring top elements.  
In the first step of the experiment we measure the S-parameters of two adjacent antennas in the absence and in the presence of the manufactured metasurface sample with $3 \times 15$ unit cells. The aim of this step is to ensure isolation improvement of the antennas due to the finite-size structure with EBG properties.

In the second step we test the same setup in an actual MR experiment to check the B1+ field distribution inside the phantom. The aim of this step is to show that the presence of the metasurface improves B1+ efficiency when one of the corresponding decoupled antennas is active. 

\subsection{Measurements of S-Parameters}

\begin{figure}
  \centering
  \includegraphics[width=0.9\linewidth]{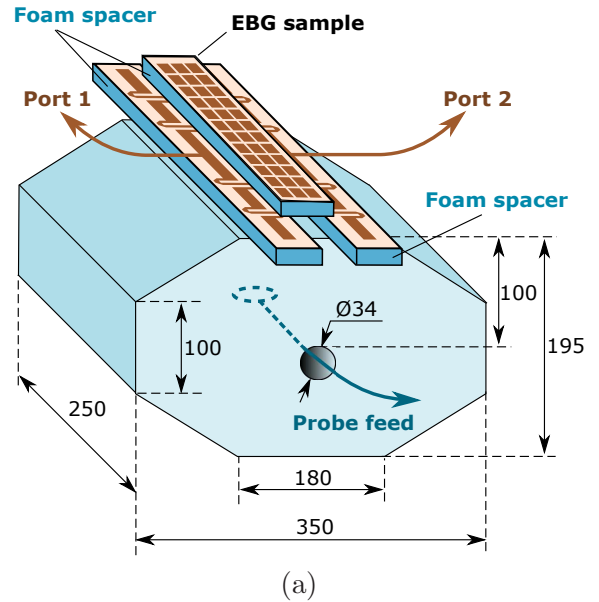}\\(a)\\
  \caption{Experimental setup for S-parameters measurements (two of six dipoles on top of the phantom are shown). Phantom dimensions are given in millimeters.}
  \label{Fig4}
\end{figure}

In order to study the decoupling capabilities of the manufactured sample with respect to the pair of top array elements we measure S-parameters by means of the Vector Network Analyzer (VNA) Planar TR 1300/1. The measurement setup is depicted in Figure \ref{Fig4}, where Dipoles 1 and 2 are equipped with the metasurface sample. The mutual positions of the antennas and the metasurface correspond to Figure \ref{FigSetup}. Also Figure \ref{Fig4} includes dimensions of the homogeneous phantom. Two 50-Ohm ports of the VNA are connected to the terminals of the dipoles. The other four dipoles on the bottom were terminated with matched loads.

The active dipoles were matched at 298 MHz by a LC lumped-element network optimized in CST Microwave Studio Schematic for the position of the antennas on the phantom inside a dummy MRI bore. The design of the matching network is based on the preparatory measured S-parameters of the setup with non-matched antennas. The matching networks are identical for both of the top antennas and were optimized in the absence and in the presence of the metasurface. For example with the metasurface each circuit contained series inductances of 10 nH and parallel capacitance of 15 pF.

\begin{figure}
  \centering
  \includegraphics[width=1\linewidth]{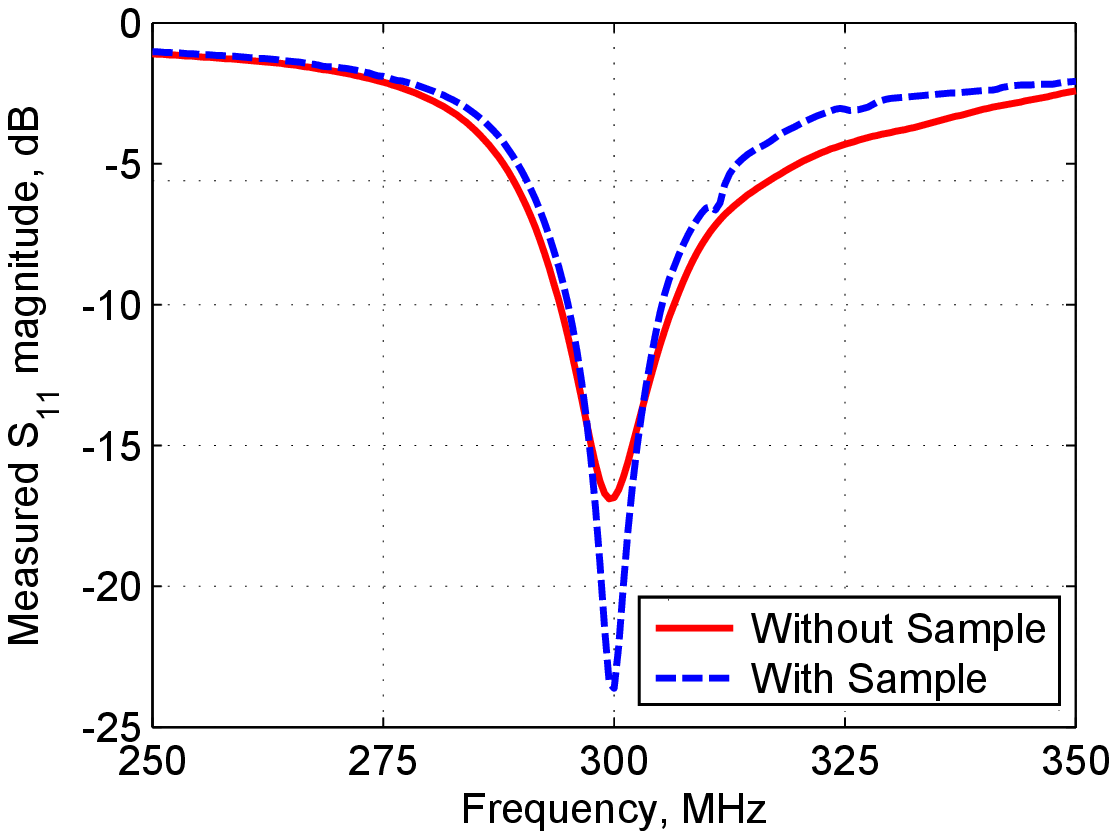}\\(a)\\
  \includegraphics[width=1\linewidth]{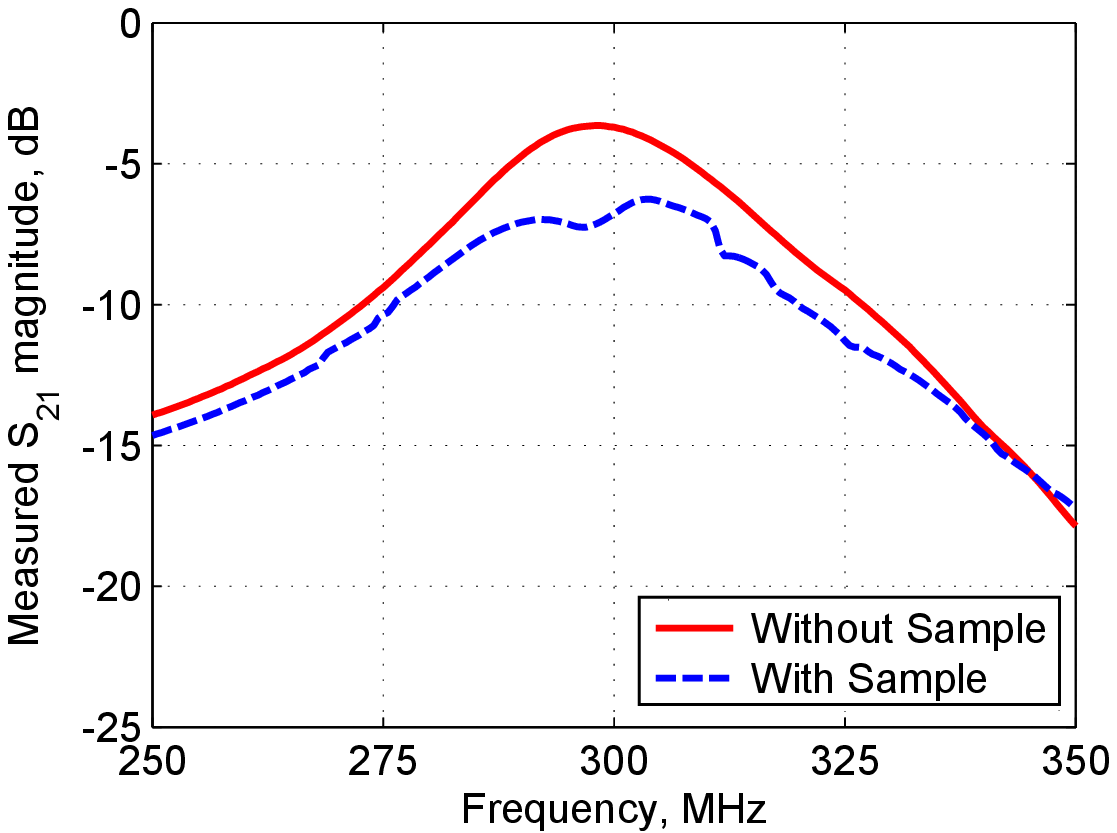}\\(b)\\
  \includegraphics[width=1\linewidth]{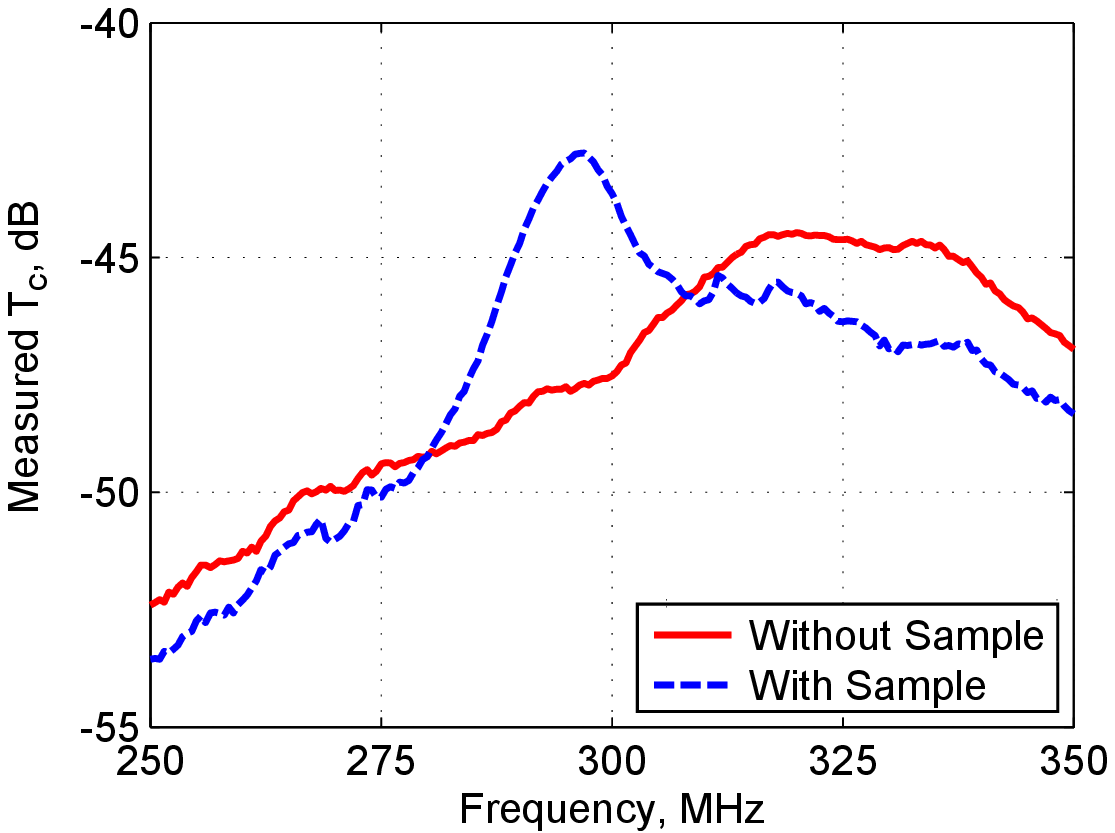}\\(c)
  \caption{(a--b) Measured S-parameters of the dipoles 1 and 2 over the phantom in the presence (dashed line) and in the absence (solid line) of the metasurface sample; (c) measured scattering coefficient $T_{\text{c}}$ for the Dipole 1 and the magnetic probe antenna at the phantom's center}
  \label{Fig5}
\end{figure}

The measured dependence of $|S_{11}| \approx |S_{22}|$ on frequency with and without the EBG metasurface is shown in Figure \ref{Fig5}(a). From this plot it is clearly seen that the reflection loss in both of the cases is below -15 dB at 298 MHz, and therefore the isolation between the top dipoles 1 and 2 is completely determined by the magnitude of $S_{21}$. The corresponding plots of $|S_{21}|$ vs. frequency are presented in Figure \ref{Fig5}(b), from which one can ensure that the presence of the metasurface indeed suppresses their mutual coupling. The level of isolation is improved by approximately -3.6 dB exactly at 298 MHz.
By means of VNA it is also possible to estimate B1+ efficiency by measuring the transmission coefficient between one active antenna and the flux probe located at the center of the phantom. For this purpose we measured the transmit efficiency parameter $T_{\text{c}}$, which is the magnitude of the non-diagonal S-matrix parameter of Dipole 1 and the flux probe located in the phantom's central hole (see Figure \ref{Fig4}). This value is proportional to the transverse magnetic field created by Dipole 1 at the center and can be used as a goal function for optimization of B1+ efficiency. When $T_{\text{c}}$ is measured, the other antennas of the array including Dipole 2 are terminated with matched loads. As the flux probe we used the Faraday's shielded loop antenna fed by a coaxial cable. The measured results are shown in Figure \ref{Fig5}(c), from which one can conclude that at 298 MHz the metasurface sample causes an improvement of the local magnetic field level at the phantom center by 4.8 dB.

\subsection{MRI measurements}

To verify that reduced coupling indeed results in increased transmit efficiency, an MR experiment was performed using the same setup as depicted in Figures \ref{FigSetup} and \ref{Fig4} on 7 Tesla MR system (Achieva, Philips Healthcare, Cleveland, USA) equipped with multi-transmit functionality (8x2 kW RF power) and bi-directional couplers on all channels. 

B1+ measurements were performed with one of the two top elements active (and all other elements passive during transmit, but participating in receive). Before each measurement, 
the coupling coefficients between the channels were determined using the bi-directional couplers and a one-time calibration procedure. 

The scattering parameters that were measured by the scanner are compared to the ones measured on the bench using the VNA in Table \ref{t1}. The comparison contains the measured parameters $S_{12}=S_{21}=T^{\text{VNA}}$ obtained on the bench by the two-port VNA at 298 MHz for the setup depicted in Figure  \ref{Fig4} (located in free space) and the two S-parameters of the 6-port setup as measured in the scanner: $|S^{\text{MRI}}_{12}|$ and $|S^{\text{MRI}}_{21}|$. The difference between these values measured in different ways is caused by slight differences in the transmit chains that were not accounted for by the calibration. 
\begin{table} 
\caption{Comparison of coupling quantities measured by VNA ($|T^{\text{VNA}}|$) and by scanner ($|S_{12}^{\text{MRI}}|$ and $|S_{21}^{\text{MRI}}|$) in presence and in absence of the metasurface sample}\label{t1}
\centering
\begin{tabular}{|m{3.2cm}|m{1cm}|m{1cm}|m{1cm}|}
\hline
Coupling Quantity, dB & $|T^{\text{VNA}}|$ & $|S_{12}^{\text{MRI}}|$ & $|S_{21}^{\text{MRI}}|$ \\
\hline
Without EBG & -3.7 & -7.3 & -6.5 \\
\hline
With EBG & -7.2 & -11.4 & -10.6 \\
\hline
Improvement & 3.5 & 4.1 & 4.1 \\
\hline
\end{tabular}
\end{table}
These results again confirm that the finite-size metasurface sample indeed provides better isolation between Dipoles 1 and 2 (even in the presence of the scanner's RF shield). It is important to mention that the reflection coefficients in the corresponding scanner/VNA channels $|S_{11}|$ and $|S_{22}|$ were below -11.5 dB for both the experiments (with and without the metasurface sample).

\begin{figure}[h!]
\center
\includegraphics[width=0.95\linewidth]{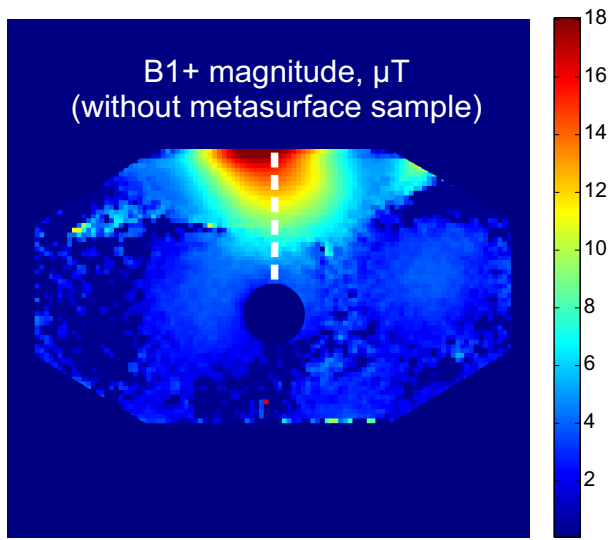}\\(a) \\
\includegraphics[width=0.95\linewidth]{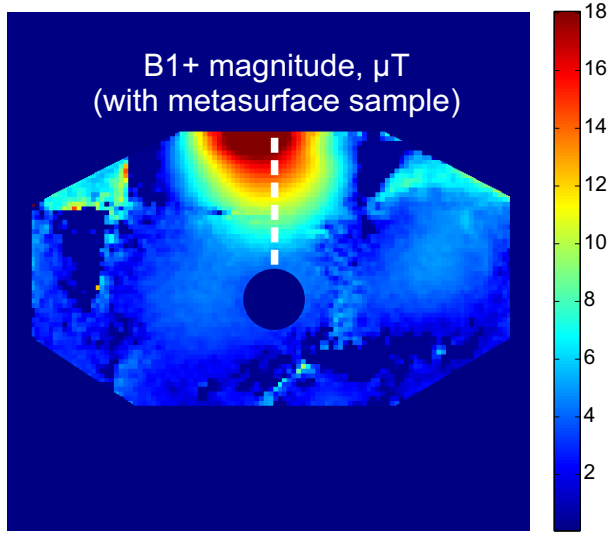}\\(b)
\caption{Measured B1+ patterns created by Dipole 1: (a) without metasurface sample; (b) with metasurface sample. Air regions (outside of phantom and inside the hole) are not shown.}
\label{Fig6}
\end{figure}

The measured B1+ distribution patterns, created by Dipole 1 in the absence (a) and in the presence (b) of the EBG metasurface are given in Figure \ref{Fig6}. Both patterns are given in the central transverse plane. The dimensions of the field of view are $376.5~\times~210$ mm in both cases. For both the cases, AFI B1+ maps were acquired \cite{Yarnykh} with TR=40/200 ms. 

\begin{figure}
  \centering
  \includegraphics[width=0.95\linewidth]{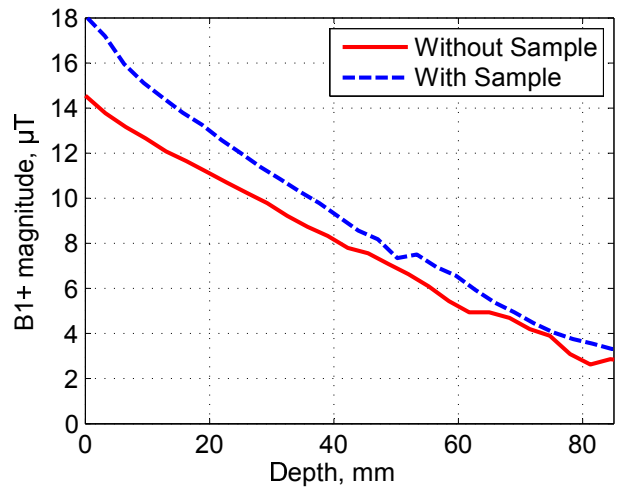}
  \caption{Measured B1+ magnitude created by Dipole 1 vs. depth of the observation point inside the phantom: with and without metasurface sample}
  \label{Fig7}
\end{figure}

In Figure \ref{Fig7} the values of B1+ magnitude created by Dipole 1 measured in the absence and in the presence of the EBG sample are compared as functions of the coordinate along the segment depicted in Figure \ref{Fig6} with a white dashed line. This comparison plot illustrates the effect of the metasurface sample on the B1+ magnitude profile from the surface to the center of the phantom (depth from 0 to 85 mm), when Dipole 1 is active. 
Figure \ref{Fig7} shows that the presence of the metasurface sample at a height of $d_3=45$ mm above Dipole 1 and Dipole 2 indeed improves the B1+ efficiency of the antennas 
demonstrating uniformly increased B1+ magnitude by up to 24\% without distortion of the initial field distribution. Therefore it is possible to find a location of the EBG sample above a pair of antennas, which provides better isolation and higher magnitude of the field inside the phantom while not affecting the penetration depth.
The setup studied in this work can lead to a multi-element body array design, in which each pair of adjacent dipole antennas is equipped with a metasurface sample. In this design the relative position of the antennas and the sample within each section of the array should remain the same as shown in Figure \ref{Fig4}.

\section{Conclusion}\label{Conclude}

In this work we first experimentally confirmed decoupling and B1+ efficiency improvement by using finite-size metasurface samples in multi-element dipole body arrays for 7T MRI. The effect was verified by vector network analyzer and flux probe measurements as well as on a 7T MR system. It was shown that the presence of the manufactured sample based on the miniaturized stacked metasurface with overlapping square patches (so called mushroom surface) at a certain electrically small distance from a pair of dipole elements homogeneously increases the B1+ magnitude inside a water phantom. We observed an isolation improvement of 3.5 dB for two adjacent dipoles of a six-element array (coupling change from -3.7 to -7.2 dB) outside of the MRI bore and 4.1 dB improvement (coupling worst-case change from -6.5 to -10.6 dB) measured directly by the MRI machine. These results were obtained for the dipoles over a phantom with separation of only 3 cm (0.03 of the wavelength in air) at 298 MHz. By MRI B1+ measurements on the scanner it was shown that the B1+ efficiency is improved due to the metasurface by up to 24\%, which can be explained by both of the following two effects. First is reduction of scattering losses due to suppression of mutual coupling between adjacent dipoles, while the second is the constructive interference between the dipole field and the field scattered by the sample. 

Based on the measured B1+ patterns and S-parameters of the setup we can confirm that finite-size samples of high-impedance metasurfaces can increase B1+ efficiency of 7T dipole-type coil-arrays, as it was theoretically proposed for single antennas in \cite{Saleh12} and explained by effective magnetic wall properties. Additionally we showed that in the array configuration such metasurface samples though having electrically small dimensions still cause considerable isolation improvement, which also increases efficiency through reduction of scattering losses. Our conclusion is that electromagnetic bandgap structures are promising tools for decoupling dipole antennas for 7T MRI. We believe that the observed effects will be further explored and find applications in future coil array designs combined with metasurfaces.

\section{Acknowledgements}

This work was supported by the Russian Science Foundation (Project No. 15-19-20054). The authors thank Prof. Constantin Simovski for useful discussion.

\bibliography{his-mri}

\end{document}